\documentclass[prb,floatfix,showpacs,superscriptaddress,twocolumn,nofootinbib]{revtex4-1}
\usepackage{amsfonts}
\usepackage{amssymb}
\usepackage{amsmath}
\usepackage{graphicx}
\usepackage{dcolumn}
\usepackage{bm}
\usepackage{epsfig}
\usepackage{color}
\usepackage[T2A]{fontenc}
\usepackage{enumitem}
\usepackage{chngcntr}

\begin{document}

\title{Damping of Confined Excitations Modes of 1D Condensates in an Optical
Lattice}
\author{C.~Trallero-Giner}
\affiliation{Facultad de F\'{\i}sica, Universidad de La Habana, Vedado 10400, La Habana,
Cuba}
\author{Dar\'{\i}o G.~Santiago-P\'{e}rez}
\affiliation{Universidad de Sancti Spiritus \textquotedblleft Jos\'{e} Mart\'{\i} P\'{e}%
rez\textquotedblright , Ave. de los M\'{a}rtires 360, CP 62100, Sanct\'{\i}
Spiritus, Cuba}
\affiliation{CLAF - Centro Latino-Americano de F\'{\i}sica, Avenida Venceslau Braz, 71,
Fundos, 22290-140, Rio de Janeiro, RJ, Brazil}
\author{Ming-Chiang Chung}
\affiliation{Department of Physics, National Chung Hsing University, Taichung, 40227, Taiwan}
\author{G. E.~Marques}
\affiliation{Departamento de F\'{\i}sica, Universidade Federal de S\~{a}o Carlos,
13565-905, S\~{a}o Carlos, Brazil}
\author{R.~Cipolatti}
\affiliation{Instituto de Matem\`{a}tica, Universidade Federal do Rio de Janeiro, C.P.\
68530, Rio de Janeiro, RJ, Brazil}
\date{\today }

\begin{abstract}
We study the damping of the collective excitations of Bose-Einstein
condensates in a harmonic trap potential loaded in an optical lattice. In
the presence of a confining potential the system is non-homogeneous and the
collective excitations are characterized by a set of discrete confined
phonon-like excitations. We derive a general convenient analytical
description for the damping rate, which takes into account, the trapping
potential and the optical lattice, for the Landau and Beliaev processes at
any temperature, $T$. At high temperature or weak spatial confinement, we
show that both mechanisms display linear dependence on $T$. In the quantum
limit, we found that the Landau damping is exponentially suppressed at low
temperatures and the total damping is independent of $T$. Our theoretical
predictions for the damping rate under thermal regime is in completely
correspondence with the experimental values reported for 1D condensate of
sodium atoms. We show that the laser intensity can tune the collision
process, allowing a \textit{resonant effect} for the condensate lifetime.
Also, we study the influence of the attractive or repulsive non-linear terms
on the decay rate of the collective excitations. A general expression of the
renormalized Goldstone frequency has been obtained as a function of the 1D
non-linear self-interaction parameter, laser intensity and temperature.
\end{abstract}

\maketitle

\section{Introduction}

The damping process plays a crucial role in the dynamic of Bose-Einstein
condensates (BEC). Phenomena such as superfluid phase transitions,~\cite
{Beliaev,Iigaya,Jin,Stringari,Onofrio,Burge,Mewes} Josephson effect,~\cite
{Cataliotti,Albiez} quantized vortices,~\cite{Madison,Feder,Hodby} Mott
insulator transition,~\cite{Greiner} among others,~\cite{Gaunt} are limited
by the finite lifetime of the collective excitations of the condensed atoms,
i.e. by the damping mechanisms and their dependence with the temperature.
After the experimental confirmation that the collective excitations are
damped,~\cite{Jin,Mewes,2Jin,Stamper} the behavior of the decay rates have
been of mayor interest in the physics of BEC. Thus, several studies and
theoretical calculations of the damping of excitations have been performed
in three-dimensional (3D)~\cite
{Beliaev,Tsuchiya,Fedichev,2Pitaevskii,Hohenberg,Szepfalusy,Liu,ming},
two-dimensional (2D)~\cite{Beliaev,Tsuchiya,ming,Natu} as well as
one-dimensional(1D) systems.~\cite{Ferlaino,Arahata,2Tsuchiya} For a better
understanding of the damping process, it becomes necessary to consider the
contribution of the parabolic confining potential. The behavior and
characteristics of the decay rate and its dependence with temperature,
differer radically if we are dealing or not with homogeneous systems. The
main assumption for homogeneous system is to consider the condensate density
constant in the all space.

In the present work we are dealing with a microscopic theory for the damping
rate of collective oscillations, specifically for a quasi-one-dimensional
(1D) condensate confined to a parabolic harmonic trap potential and loaded
into an optical lattice. The existence of a non-negligible confining
external potential breaks the invariance symmetry, which leads to the
damping rate showing a different qualitative behavior in comparison with
previous formalism, where the condensate is tackled as a homogeneous system
(see for example Refs.~[\onlinecite
{ming,Pitaevskii}] and references there in). The influence of both
external interactions -the trap potential and laser intensity- must provide
a physically richer scenario for the decay of collective excitations. The
knowledge of excited states or Goldstone modes enables the characterization
of the condensate dynamics in a general framework. In the case we are in
presence of spatially non-homogeneous BEC system, the label spacing of the
discrete spectrum (confined phonon-like modes) and also, the nature and
symmetry of wavefunction of the collective modes, are required for the
calculation of the collision scattering process.~\cite{Fedichev}

Cigar-shaped traps can be considered as quasi-one-dimensional systems. We
select such platform of a condensate loaded simultaneously into a 1D
harmonic potential and an optical lattice to characterize the phenomenon of
damping and tackle the problem analytically. Within the framework of mean
field theory, the physical characteristics of a BEC in such trapping profile
are ruled by the time dependent non-linear Gross-Pitaevskii equation (GPE)~
\cite{Pitaevskii,trallero} in an external potential
\begin{equation}
V_{ext}(x)=\frac{1}{2}m\omega _{0}^{2}x^{2}-V_{L}\cos ^{2}\left( \dfrac{2\pi
}{d}x\right) ,  \label{potent}
\end{equation}
where $m$ is the alkaline atom mass, $V_{L}$ the laser intensity, $d$ its
laser wavelength, and $\omega _{0}$ the frequency of the harmonic trap.

\section{Theoretical background}

\begin{figure}[tbp]
\includegraphics[width = 0.48\textwidth]{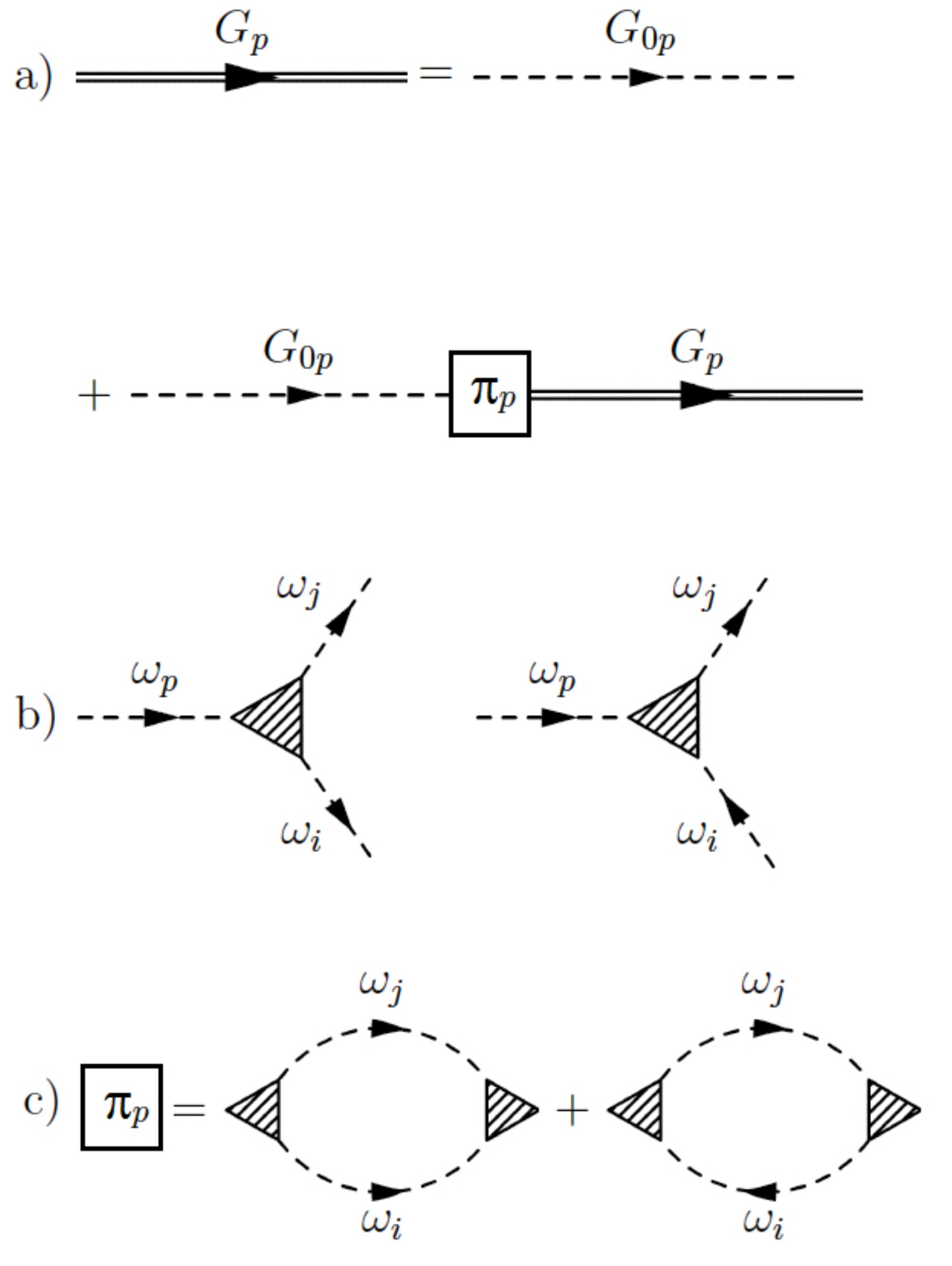}
\caption{a) Diagrammatic equation for casual Green function $G_{p}$ of
phonon modes in the condensate. b) Vertices involved in the collision
phonon-phonon cubic interaction. c) Feynman diagrams contributing to phonon
self-energy operator $\protect\pi _{p}$ for Beliaev and Landau processes.}
\label{feyman_diagram}
\end{figure}

In the framework of the Green function formalism, the spectrum of the
excited states is obtained by the poles of the dressed Green function $G_{p}$
. The solution of the Dyson equation, shown diagrammatically in Fig.~\ref
{feyman_diagram}a), is the renormalized Green function $G_{p}$ given by
\begin{equation}
G_{p}^{-1}=G_{0p}^{-1}-\pi _{p}~.  \label{Secular_equation}
\end{equation}%
In absence of interaction $G_{p}\rightarrow G_{0p}=[\omega -\omega
_{p}+i\varepsilon ]^{-1}$, $\varepsilon >0,$ $\omega _{p}$ is the
eigenfrequency of the excited state, and $\pi _{p}$ is the self-energy
contribution.
The solution of Eq.~(\ref{Secular_equation}) leads to the complex frequency $
\omega =\omega _{p}+\pi _{p}$. Here, \emph{Re}($\pi _{p}$) represents the
renormalized contribution to the eigenfrequency $\omega _{p}$, while \emph{\
Im }($-\pi _{p}$) corresponds to the damping rate.

We assume that the decay processes are associated to the collision between
confined phonon states. In first order, the collision term is described by
the interaction between three interacting phonon modes, giving rise to the
vertices in the self-energy diagrams shown in Fig.~\ref{feyman_diagram}b).
It becomes clear that the last diagram in Fig~\ref{feyman_diagram}b) does
not contribute to the self-energy interaction at $T=0$ K, since a thermal
excited mode $\omega _{i}$ must be present in the system (see Eq.~(\ref
{omega}) below). We must recall that the thermal cloud in the present theory
is assumed in thermal equilibrium.

Figure ~\ref{feyman_diagram}c) presents the leading diagrams contributing to
the self-energy $\pi _{p}$. Accordingly, the complex frequency correction
can be factorized into two main processes as $\pi _{p}=\Delta \tilde{\omega}
_{L}+\Delta \tilde{\omega}_{B}$ and, in the Hartree-Fock-Bogoliubov
approximation we obtain that~\cite{ming}
\begin{eqnarray}
\pi _{p} &=&\frac{2\pi g_{1}^{2}}{\hbar ^{2}}\sum\limits_{i,j}\left[ \frac{
2\left( f_{i}-f_{j}\right) \left\vert A_{ij}\right\vert ^{2}}{\omega
_{p}+\omega _{i}-\omega _{j}+i\varepsilon }+\right.  \notag \\
&&\left. \frac{\left( 1+f_{i}+f_{j}\right) \left\vert B_{ij}\right\vert ^{2}
}{\omega _{p}-\omega _{i}-\omega _{j}+i\varepsilon }\right] ~,  \label{omega}
\end{eqnarray}
where $g_{1}$ is the coupling constant, $A_{ij}$ ($B_{ij}$) represents the
matrix elements for the Landau (Beliaev) process $\omega _{p}+\omega
_{i}\rightarrow \omega _{j}$ ($\omega _{p}\rightarrow $ $\omega _{i}+\omega
_{j}$). In Eq.~(\ref{omega}) the sum $\sum\limits_{i,j}$ takes into account
all possible virtual transitions $\left\vert\omega _{i}\right\rangle $ and $
\left\vert \omega _{j}\right\rangle,$, contributing to the decay rate, while
the term $1+f_{i}+f_{j}$ gives us the Bose-Einstein statistical factor of
the phonon $\left\vert \omega _{p}\right\rangle $ decaying into two confined
phonon modes (first diagram in Fig.~\ref{feyman_diagram}c)) $\left\vert
\omega _{i}\right\rangle $ and $\left\vert \omega _{j}\right\rangle ,$ and $
f_{i}-f_{j}$ corresponds to the thermal correction for the annihilation and
creation of phonons with frequencies $\omega _{i}$ and $\omega _{j}$ (second
diagram in Fig.~\ref{feyman_diagram}c))$,$ respectively.

\subsection{Bogoliubov type excitations}

Bogoliubov type excitations can be searched by applying a small deviation
from the GPE stationary solutions $\left\vert \psi _{0}\right\rangle \exp
(-i\mu t/\hbar )$, i.e.
\begin{multline}
\left\vert \Psi (t)\right\rangle =\exp (-i\mu t/\hbar )\left[ \left\vert
\psi _{0}\right\rangle +\left\vert u\right\rangle \exp (-i\omega t)\right. \\
\left. +\left\vert v^{\ast }\right\rangle \exp (i\omega t)\right] \text{ },
\label{eq:wafeper}
\end{multline}
with $\mu $ being the chemical potential. By linearizing the time-dependent
non-linear GPE, we obtain the Bogoliubov-de Gennes equations (B-dGE) for the
eigenfrequencies $\omega $ and amplitudes $\left\vert u\right\rangle $ and $
\left\vert v\right\rangle.$ As stated above, if the harmonic trap potential
is switched off, we are in presence of the homogeneous case, where the
phonon wavevector $\mathbf{q}=q_{x}\mathbf{e}_{x}$ is a good quantum number
and with excited frequency $\omega (q_{x})$ depending on the phonon
wavevector. Thus, the Bogoliubov's low-lying excitation spectrum shows a
linear phonon dispersion in $q_{x}$. On the other hand when the condensate
is loaded into a harmonic potential, $\omega _{0}\neq 0$ in Eq.~(\ref{potent}
), the system becomes inhomogeneous and the wavevector $\mathbf{q}$ is no
longer a good quantum number. In such case the B-dGE provides a set of
discrete excited state frequencies $\omega _{p}$ ($p=1,2,...$).

By considering the periodic potential in (\ref{potent}) and the non-linear
term, $g_{1}\left\vert \phi _{0}\right\vert ^{2}$, in the GPE as a
perturbation with respect to the harmonic trap potential, $\frac{1}{2}
m\omega _{0}^{2}x^{2}$, phonon frequencies $\omega _{p}$ up to the second
order in $g_{1}$ are given by~\cite{trallero2}
\begin{multline}
\frac{\omega _{p}}{\omega _{0}}=p+\frac{\Lambda }{\sqrt{2\pi }}\left[ -1+
\frac{2\Gamma (p+1/2)}{\sqrt{\pi }p!}\right] -  \label{frecue} \\
\left. \frac{V_{0}}{2}\exp \left( -\alpha ^{2}\right) \times \left[
L_{t}(2\alpha ^{2})-1\right] -\right. \\
\left. \frac{\Lambda V_{0}}{\sqrt{2\pi }}\exp \left( -\alpha ^{2}\right)
\left[ Ei(\frac{\alpha ^{2}}{2})-\mathcal{C}-\ln \frac{\alpha ^{2}}{2}+\frac{
\delta _{p}(\alpha )}{\sqrt{\pi }}\right] +\right. \\
\left. \frac{V_{0}^{2}}{4}\exp \left( -2\alpha ^{2}\right) \left[
Chi(2\alpha ^{2})-\mathcal{C-}\ln 2\alpha ^{2}+\rho _{p}(\alpha )\right]
+\right. \\
+\Lambda ^{2}\left[ \frac{\gamma _{p}}{2\pi ^{2}}+0.033106\right] \text{ ,\
\ \ \ \ }p=1,2,....,
\end{multline}%
with $\Lambda =g_{1}N/(l_{0}\hbar \omega _{0})$, $N$ the number of atoms, $
l_{0}=\sqrt{\hbar /m\omega _{0}}$ defining the characteristic unit length, $
\alpha =2\pi l_{0}/d$ and $V_{0}=V_{L}/\hbar \omega _{0}.$ In addition $
L_{t}(z),$ $\Gamma (z)$, $Ei(z), $ $Chi(z)$, and $\mathcal{C}$ are the
Laguerre polynomials, the gamma function, the exponential integral, the
cosine hyperbolic integral and the Euler's constant, respectively. The
parameters $\gamma _{p}$, $\delta _{p}$ and $\rho _{p}$ are reported
elsewhere.~\cite{trallero2}

\subsection{Symmetry of the excited states}

Owing to the inversion symmetry, the space of solutions can be decoupled
into two independent subspaces, $\mathcal{O}$ and $\mathcal{E}$ for $
p=1,3,....$ and $p=2,4,....$ modes, respectively. Hence, the components $
\left\vert u_{p}\right\rangle $ and $\left\vert v_{p}\right\rangle $ are
expanded over the complete set 1D oscillator wave functions $\left\{ \phi
_{2p+1}\right\} $ or $\left\{ \phi _{2p}\right\} $ for the odd, $\mathcal{O}$
and even, $\mathcal{E}$ Hilbert subspaces. The normalized eigenvectors $
\left\vert \Phi _{p}\right\rangle ^{\dagger }=\left[ \left\vert u_{p}^{\ast
}\right\rangle ,\left\vert v_{p}\right\rangle \right] ,$ up to first order
in $\Lambda $ and $V_{0},$ can be cast as~\cite{trallero2}
\begin{equation}
\left\vert \Phi _{p}\right\rangle =\left(
\begin{array}{c}
\left\vert \phi _{p}\right\rangle +\sum\limits_{m\neq p}\dfrac{\left(
4\Lambda f_{p,m}-V_{0}g_{p,m}\right) }{2(p-m)}\left\vert \phi
_{m}\right\rangle \\
\\
-\sum\limits_{m=0}^{\infty }\dfrac{\Lambda f_{p,m}}{p+m}\left\vert \phi
_{m}\right\rangle
\end{array}
\right) \text{ },  \label{Fik}
\end{equation}
with
\begin{equation}
f_{p,m}=\dfrac{\left( -1\right) ^{(p-m)/2}}{\pi \sqrt{2m!p!}}\Gamma \left(
\frac{p+m+1}{2}\right) \text{ },  \label{f}
\end{equation}
\begin{multline}
g_{p,m}=\frac{\left( -1\right) ^{(p-m)/2}}{\sqrt{m!p!}}h!\left( 2\alpha
^{2}\right) ^{(p-m)/2}\times  \label{g} \\
\exp \left( -\alpha ^{2}\right) L_{h}^{\left\vert p-m\right\vert }(2\alpha
^{2})\text{ ,}
\end{multline}
$L_{h}^{t}(2\alpha ^{2})$ the Associate Laguerre polynomials, $
h=(p+m-\left\vert p-m\right\vert )/2$ and $m+p$ is an even number.

The parity of the function $\left\vert \Phi _{p}\right\rangle $ is linked to
the index $p,$ if $p$ is even or odd the eigenstate $\left\vert \Phi
_{p}\right\rangle $ is symmetric or antisymmetric. The decay process of a
certain phonon $p$ is restricted by the symmetry property of the matrix
elements in Eq.~(\ref{omega}). The amplitudes $A_{ij}(p)$ and $B_{ij}(p)$
impose a parity selection rule for the involved states $\left\vert \Phi
_{p}\right\rangle ,$ $\left\vert u_{i}\right\rangle $ and $\left\vert
v_{i}\right\rangle $. As shown in Eqs.~(\ref{Bij}) and (\ref{A}), for a
symmetric (antisymmetric) state $\left\vert \Phi _{p}\right\rangle ,$ the
amplitudes $\left\vert u_{i}\right\rangle $ and $\left\vert
v_{i}\right\rangle $ must fulfill the parity condition $i+j=$ even (odd)
number, therefore limiting the possible number process for Beliaev, $\omega
_{p}\rightarrow \omega _{i}+\omega _{j},$ and Landau, $\omega _{p}+\omega
_{i}\rightarrow \omega _{j}$ decay rates. Besides the symmetry of the matrix
elements $A_{ij}(p)$ and $B_{ij}(p),$ for certain eigenmode $\left\vert \Phi
_{p}\right\rangle $ with frequency $\omega _{p}$, a key role in the damping
process is ruled by the label spacing between the Bogoliubov collective
oscillations $\Delta _{p}^{(i,j)}=\left( \omega _{p}-\omega _{i}-\omega
_{j}\right) /\omega _{0}$ and $\Delta _{j}^{(i,p)}=\left( \omega _{j}-\omega
_{i}-\omega _{p}\right) /\omega _{0}.$ Fixing the frequency $\omega _{p},$
all allowed combinations of $\omega _{i}$ and $\omega _{j}$ approaching $
\Delta _{p}^{(i,j)}$ or $\Delta _{j}^{(i,p)}$ to zero value, lead to
resonant transitions for the Beliaev or Landau damping processes.
\begin{figure}[tbp]
\includegraphics[width = 0.48\textwidth]{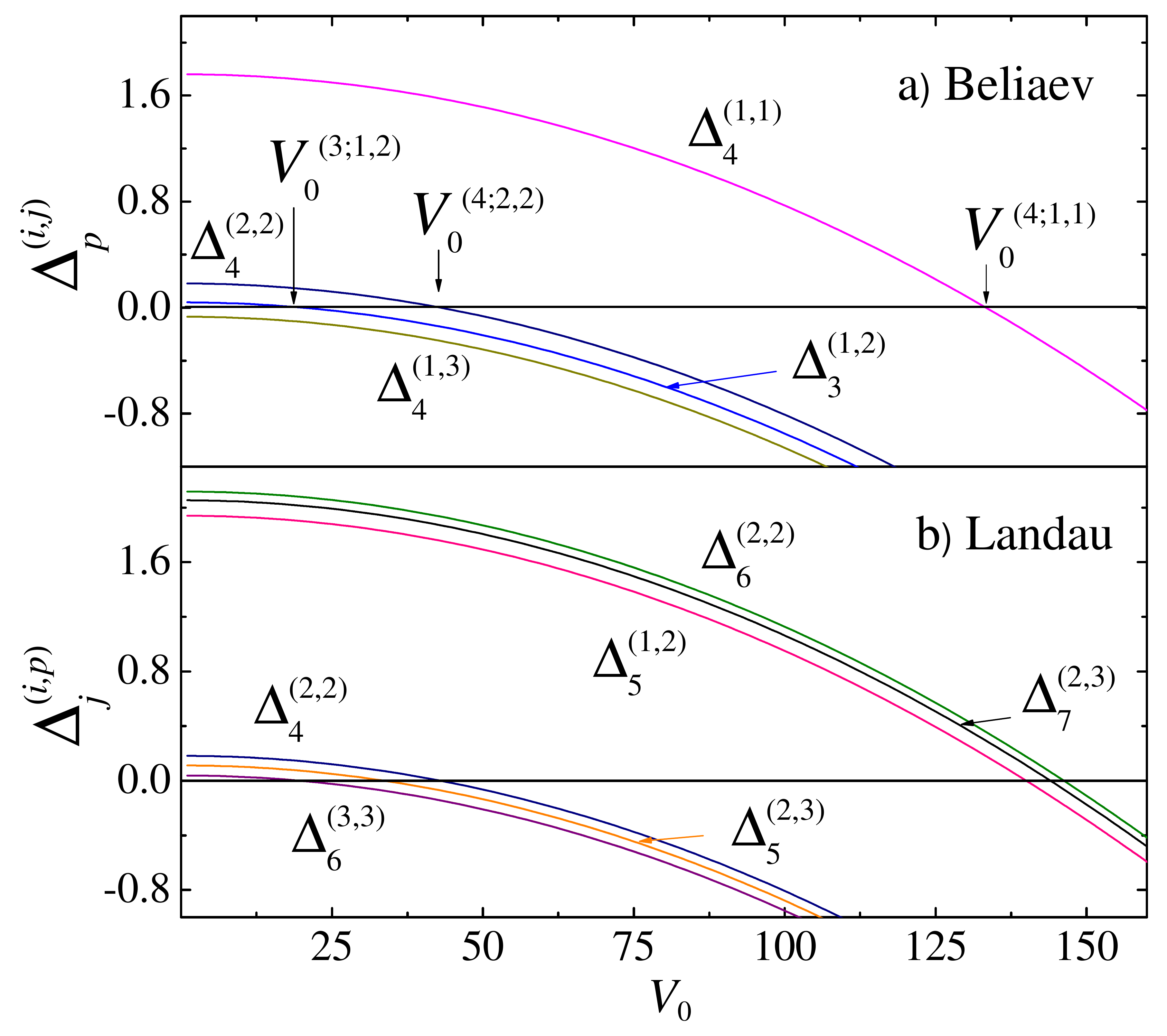}
\caption{(Color online) Dependence of the label spacing $\Delta
_{p}^{(i,j)}=\left( \protect\omega _{p}-\protect\omega _{i}-\protect\omega
_{j}\right) /\protect\omega _{0}$ and $\Delta _{j}^{(i,p)}=\left( \protect
\omega _{j}-\protect\omega _{i}-\protect\omega _{p}\right) /\protect\omega
_{0}$ (see Eq.~(\protect\ref{omega})) on the reduced laser intensity $V_{0}$
for Beliaev and Landau damping rates, respectively. Critical values, $
V_{0}^{(p;i,j)}$, where $\Delta _{p}^{(i,j)}$ approaches zero, are shown by
arrows.}
\label{fig:figD}
\end{figure}
This effect is shown in Fig.~\ref{fig:figD}, where the influence of the
intensity $V_{0}$ on some level spacing $\Delta _{p}^{(i,j)}$ (panel a)) and
$\Delta _{j}^{(i,p)}$ (panel b)) are displayed. For calculations we fixed $
d/l_{0}=0.25$ and $\Lambda =2.$ From the figure, it can be seen that the
laser intensity can be used as a external parameter to tune particular
damping process, i.e. we are able to reach certain critical values $
V_{0}^{(p;i,j)}$ (see Fig.~\ref{fig:figD} a)), where $\Delta
_{p}^{(i,j)}(V_{0})=0$. This is a direct consequence of the fact that the
Bogoliubov-type collective excitation energies, as a function of $\Lambda $
and $V_{0},$ are not equidistant. Notice that, for a certain state $p$, the
Landau mechanism allows more combinations fulfilling the condition $\Delta
_{j}^{(i,p)}(V_{0})=0$.

\section{Decay rate}

In the sequel, we consider that the damping is originated by a collision
process, and in first-order approximation, it is described by the
interaction of the three confined phonons, giving rise to a cubic
interaction in the bare phonon amplitude.~\cite
{2Pitaevskii,Fedichev,Pethick,Tsuchiya,Mazets} This mechanism is represented
by the vertex diagrams of the self-energy part shown in Fig.~\ref
{feyman_diagram}b).

\subsection{Beliaev damping}

In first-order loop approximation, the Beliaev mechanism arises from the
collision of three particles where one phonon with frequency $\omega _{p}$
is annihilated decaying into two confined excitations $\omega _{i}$ and $
\omega _{j}.$ Therefore, the allowed processes for the confined modes $
\omega _{p}$ are those with $p=2,3,...$. Following Feynman diagrams of Fig.~
\ref{feyman_diagram}b) and using the eigenfunction amplitudes given in Eq.~(
\ref{Fik}), we find that the decay amplitude $B_{ij}$ can be cast as~\cite
{ming,Pitaevskii}
\begin{figure}[tbph]
\includegraphics[width = 0.48\textwidth]{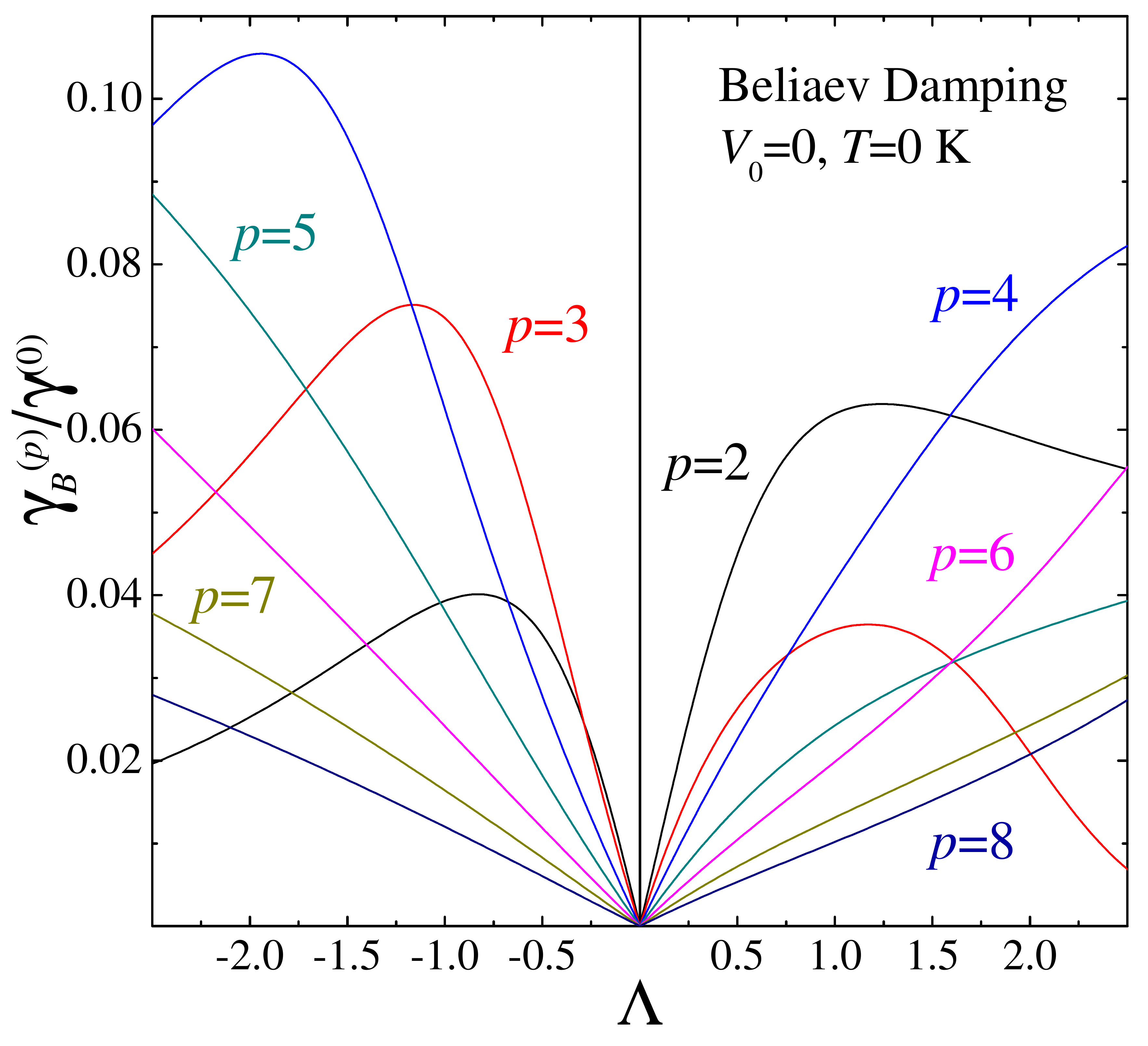}
\caption{(Color online) Reduced Beliaev damping $\protect\gamma _{B}^{(p)}/
\protect\gamma^{(0)}$ for the Goldstone modes p=2,...,8 versus the
dimensionless self-interaction parameter $\Lambda $ at laser intensity $
V_{0}=0$ and $T=$0 K.}
\label{fig:fig1}
\end{figure}
\begin{multline}
B_{ij}=\int dx\psi _{0}\left[ u_{p}\left( u_{i}^{\ast }u_{j}^{\ast
}+u_{i}^{\ast }v_{j}^{\ast }+v_{i}^{\ast }u_{j}^{\ast }\right) +\right.
\label{Bij} \\
\left. v_{p}\left( u_{i}^{\ast }v_{j}^{\ast }+v_{i}^{\ast }v_{j}^{\ast
}+v_{i}^{\ast }u_{j}^{\ast }\right) \right] \text{ }.
\end{multline}
Thus, for the Beliaev damping rate we obtain
\begin{equation}
\gamma _{B}^{(p)}=\frac{\gamma ^{(0)}}{2}\left\vert \Lambda \right\vert
\mathcal{M}_{p}^{(B)}(\Lambda ,V_{0})\text{ },  \label{Beli}
\end{equation}
with $\gamma ^{(0)}=4\pi g_{1}/(l_{0}\hbar)$ and $\mathcal{M}
_{p}^{(B)}(\Lambda ,V_{0})$ being defined in the Appendix A~. Notice that
Beliaev mechanism is forbidden if $\epsilon\rightarrow 0$ in Eq.~(\ref{omega}
). The energy conservation limits the real phonon transitions $\omega _{p}
\rightarrow \omega _{i} + \omega _{j}.$

\begin{figure}[tbp]
\includegraphics[width = 0.48\textwidth]{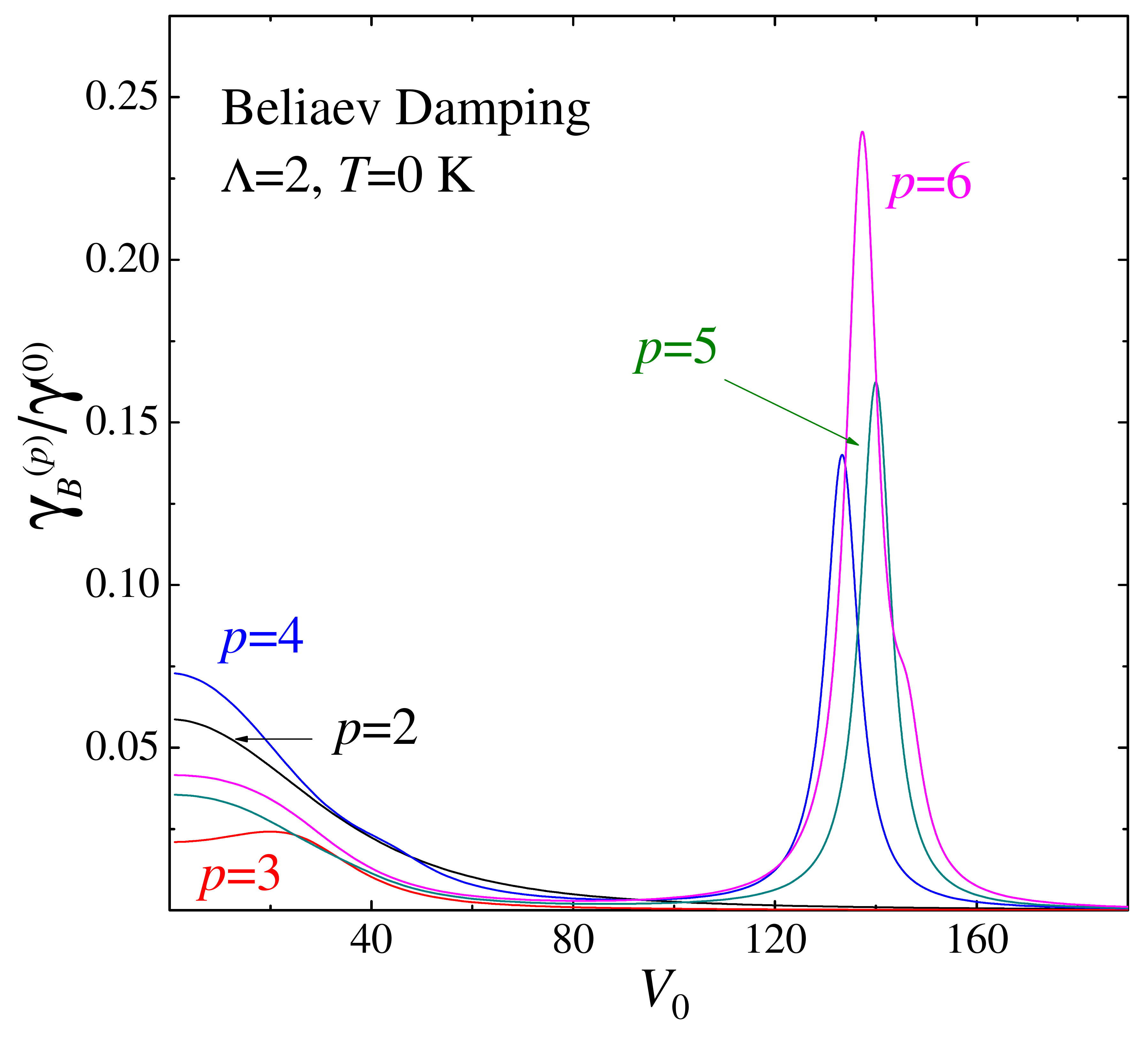}
\caption{(Color online) Influence of the laser intensity $V_{0}$ on the
reduced Beliaev damping $\protect\gamma _{B}^{(p)}/\protect\gamma^{(0)}$.
Resonant peaks are related to the zeros of the label spacing $\Delta
_{p}^{(i,j)}(V_{0})$ in Eq.~(\protect\ref{omega}).}
\label{fig:fig3}
\end{figure}

Figure~\ref{fig:fig1} displays the behavior of the Beliaev damping rate, $
\gamma _{B}^{(p)},$ in units of $\gamma ^{(0)},$ as a function of the
reduced parameter $\Lambda $ for the first seven allowed confined modes $
p=2,...,8.$ In this calculation we used $T=0$ and laser intensity $V_{L}=0$.
For small values of $\Lambda ,$ all the normalized damping seen in Fig.~\ref
{fig:fig1}, presents a linear behavior, while for increasing values of $
\left\vert \Lambda \right\vert $, the function $\gamma _{B}^{(p)}/\gamma
^{(0)}$ behaves non-monotonically and reaching a maximum. For a given
excited state $\left\vert \Phi _{p}\right\rangle ,$ the position of the
maximum is not symmetric with respect of the type of non-linear interaction
(repulsive $g_{1}>0$ or attractive $g_{1}<0$). In Fig.~\ref{fig:fig3} we
show the dependence of $\gamma _{B}^{(p)}$ on the dimensionless laser
intensity $V_{0},$ and calculated for $\Lambda =2,$ $d/l_{0}=0.25$ and $T=0$
K. It can be observed that the excited states $p=4,5$ and $6$.show sharp
peaks at certain values of $V_{0}$ These features are linked to the zeros of
the frequency label spacing
$\Delta _{p}^{(i,j)}(V_{0})$ as represented in Fig.~\ref{fig:figD}, while
the number of transitions $\omega _{p}\rightarrow \omega _{i}+\omega _{j}$
and the strength of the matrix elements, $\overline{B_{ij}},$ dictate the
relative intensity of the peaks.

\subsection{Landau damping}

Here, in the damping process a phonon mode with frequency $\omega _{p}$ and
a thermal excitation $\omega _{i}$ are annihilated and confined phonon is
created. Thus, the Landau mechanism is a thermal process at finite
temperature. In present case the vertices phonon-phonon interaction (see
Fig.~\ref{feyman_diagram}b)), conduces to the amplitude probability, \cite
{ming,2Pitaevskii}
\begin{multline}
A_{ij}=\int dx\phi _{0}\left[ u_{p}\left( u_{i}u_{j}^{\ast
}+v_{i}v_{j}^{\ast }+v_{i}u_{j}^{\ast }\right) +\right.  \label{A} \\
\left. v_{p}\left( u_{i}u_{j}^{\ast }+v_{i}v_{j}^{\ast }+v_{i}u_{j}\right)
\right] \text{ }.
\end{multline}
Thus, we have
\begin{equation}
\gamma _{L}^{(p)}=\gamma ^{(0)}\left\vert \Lambda \right\vert \mathcal{M}
_{p}^{(L)}(\Lambda ,V_{0})\text{ },  \label{Landau}
\end{equation}
where $\mathcal{M}_{p}^{(L)}(\Lambda ,V_{0})$ is defined in the Appendix B.
Hence, the total damping can be cast as
\begin{equation}
\gamma ^{(p)}=\gamma ^{(0)}\left\vert \Lambda \right\vert \left( \mathcal{M}
_{p}^{(L)}+\frac{1}{2}\mathcal{M}_{p}^{(B)}\right) \text{ }.  \label{gammap}
\end{equation}
\begin{figure}[tbph]
\includegraphics[width = 0.48\textwidth]{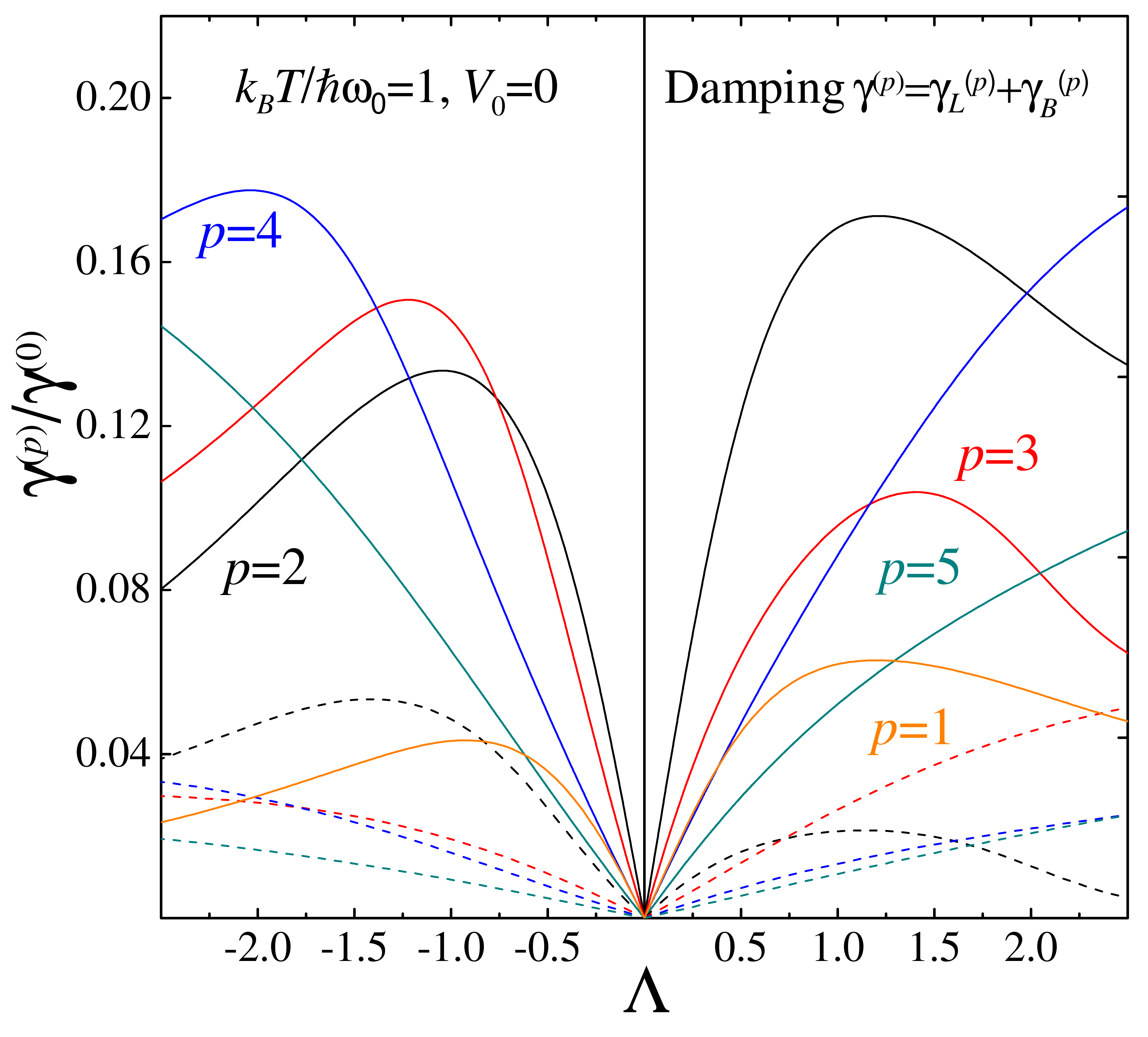}
\caption{(Color online) Total damping $\protect\gamma ^{(p)}=\protect\gamma
_{L}^{(p)}+\protect\gamma _{B}^{(p)}$ in units of $\protect\gamma^{(0)}$ for
the Goldstone modes $p=1,2,...,8$ versus the dimensionless self-interaction
parameter $\Lambda $. Landau damping for the modes are shown for equivalent
color dashed lines.}
\label{fig:fig5N}
\end{figure}
Figure~\ref{fig:fig5N} presents the total damping $\gamma ^{(p)}$ (solid
lines) as a function of $\Lambda $ for the first fifth excited states. The
Landau contribution $\gamma _{L}^{(p)}$ is represented \ by dashed lines. As
in the case of the Beliaev process, $\gamma ^{(p)}/\gamma ^{(0)}\sim
\left\vert \Lambda \right\vert $ for small values of the self-interaction
atom-atom parameter, while for large values of $\left\vert \Lambda
\right\vert $, the function $\gamma ^{(p)}/\gamma ^{(0)}$ has a maximum at
certain $\Lambda _{p}$ value. We note that $\gamma _{L}^{(p)}$ is smaller
than $\gamma _{B}^{(p)}$ for all excited states $p=2,...,5.$ For the Beliaev
damping, the first excited state $p=1$ is forbidden at any temperature,
while for $T\neq 0$ K this mode becomes allowed for the Landau process.

\begin{figure}[tbph]
\includegraphics[width = 0.48\textwidth]{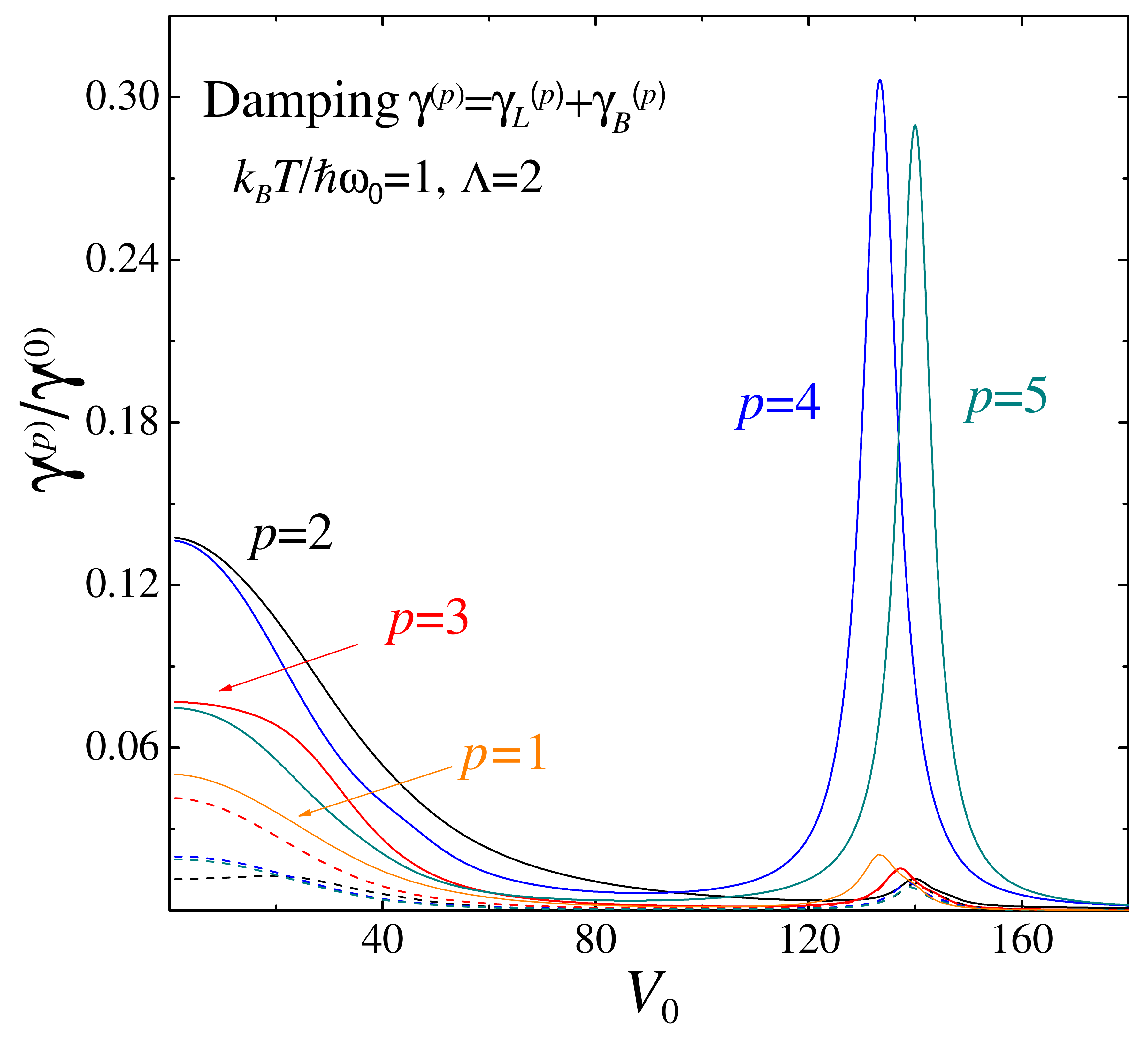}
\caption{(Color online) Influence of the laser intensity $V_{0}$ on the
total damping $\protect\gamma ^{(p)}=\protect\gamma _{L}^{(p)}+\protect
\gamma _{B}^{(p)}$ in units of $\protect\gamma^{(0)}$ for the confined modes
$p=1,...,5$. Dashed lines: Landau damping. In the calculation $d/l_{0}=0.25$
. }
\label{fig:fig6}
\end{figure}
The dependence of $\gamma ^{(p)}$ (solid lines) for $p=1,...,5$ on the laser
intensity is shown in Fig.~\ref{fig:fig6}. For sake of comparison the Landau
damping contribution is represented by dashed lines. In the figure it is
observed that the total damping presents the same behavior as the Beliaev
decay (see Fig.~\ref{fig:fig3}), also, that $\gamma _{L}^{(p)}$ shows
resonant transitions for $V_{0}\sim 140.$

\begin{figure}[tbp]
\includegraphics[width = 0.48\textwidth]{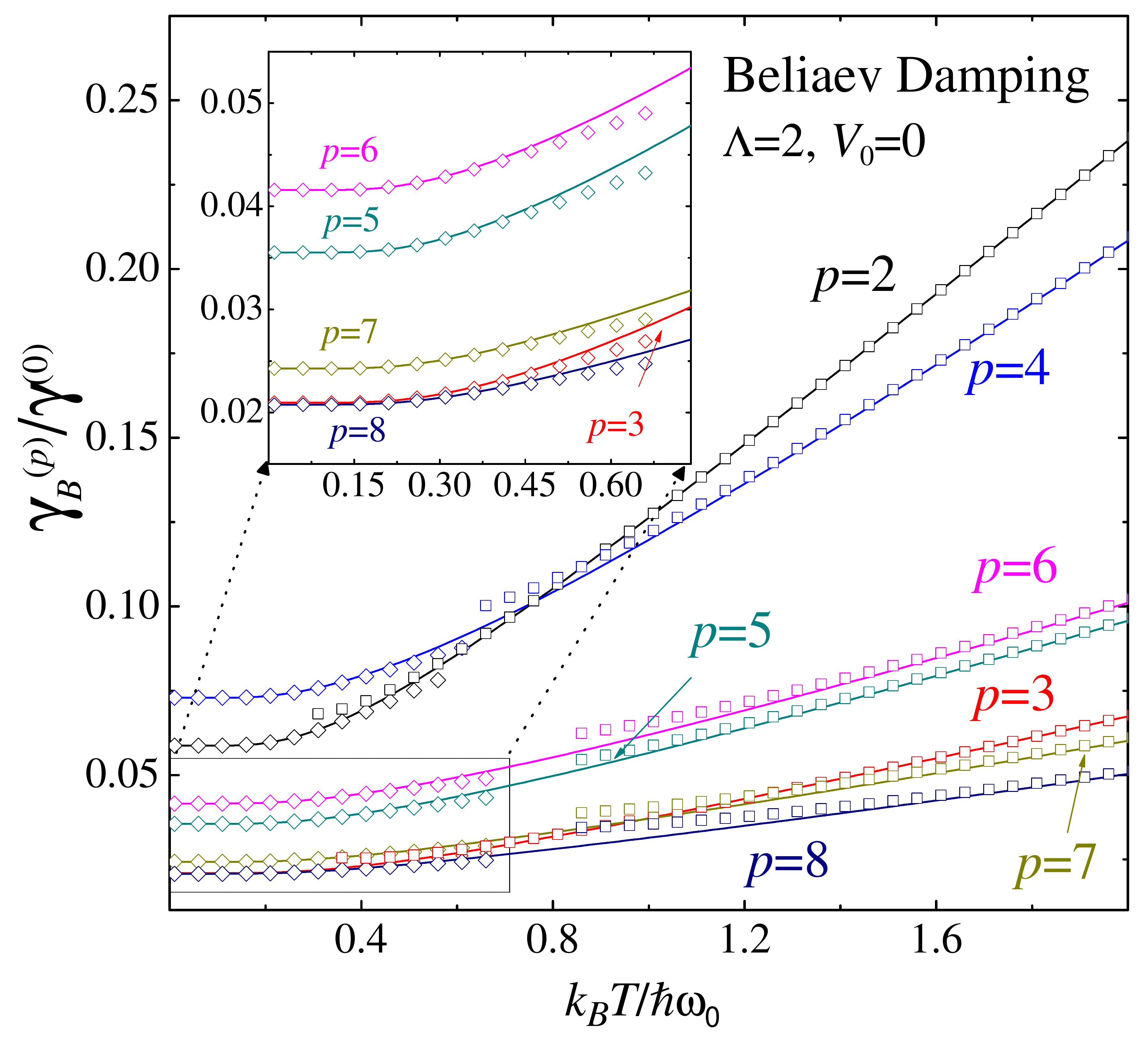}
\caption{(Color online) Reduced Beliaev decay rate $\protect\gamma
_{B}^{(p)}/\protect\gamma^{(0)}$ versus $k_{B}T/\hbar\protect\omega_{0}$ for
the Goldstone modes $p=2,...,8$. Open squares represent the solution for the
thermal regime as derived from Eq.~(\protect\ref{BETher}). Open diamonds
correspond to the low temperature limit according to Eq.~(\protect\ref
{LimTlow}).}
\label{fig:fig2}
\end{figure}

\section{Discussion of the results and conclusions}

\begin{figure}[tbp]
\includegraphics[width = 0.48\textwidth]{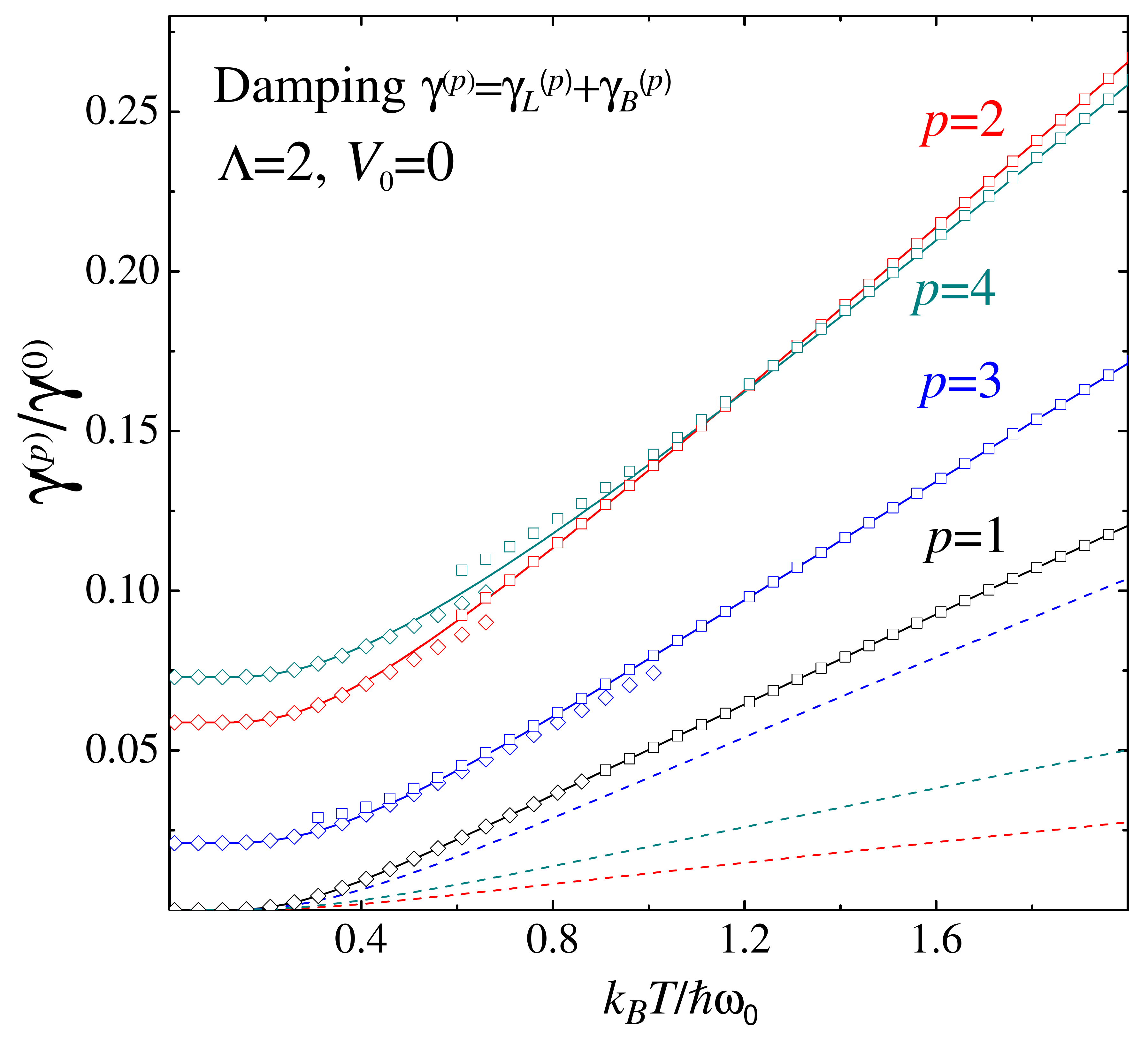}
\caption{(Color online) Reduced total damping $\protect\gamma ^{(p)}/\protect
\gamma^{(0)}$ for the Goldstone modes, $p=1,2,3,4$ as a function of
temperature. Dashed lines: Landau damping. Weak, $k_{B}T>>\hbar \protect
\omega _{0}$, and strong confinement, $k_{B}T<<\hbar \protect\omega _{0},$
limits are represented by open squares and open diamonds, respectively (see
text).}
\label{fig:fig5}
\end{figure}

From the reported calculations, two main results can be easily deduced: a)
The behavior of the damping rates $\gamma _{B}^{(p)},$ $\gamma _{L}^{(p)}$
and $\gamma ^{(p)}$ with the temperature, and b) the evaluation of the
renormalized confined phonon frequencies. Equations~(\ref{Beli}) and (\ref
{Landau}) allow a good approach for all temperature regime. Figures~\ref
{fig:fig2} and \ref{fig:fig5} depict $\gamma _{B}^{(p)}$ and $\gamma ^{(p)}$
decays for $\Lambda =2$ and $V_{0}=0$ as a function of the reduced
temperature $k_{B}T/\hslash \omega _{0}$, respectively. From Fig.~\ref
{fig:fig5} we have that, in the range of temperature here considered, $
\gamma _{B}^{(p)}$ $>$ $\gamma ^{(p)}$ for all excited states $p>1.$

In the thermal regime where $k_{B}T>>\hbar \omega _{0},$ i.e. high
temperature or weak quantum confinement, we obtain that $\gamma ^{(p)}$,
given by Eq.~(\ref{gammap}), is reduced to
\begin{multline}
\gamma ^{(p)}=\gamma ^{(0)}\left\vert \Lambda \right\vert \frac{k_{B}T}{
\hbar \omega _{0}}\left[ \mathcal{A}_{p}^{(1)}+\frac{1}{2}\mathcal{B}
_{p}^{(1)}-\right.  \label{BETher} \\
\left. \frac{1}{12}\left( \frac{\hbar \omega _{0}}{k_{B}T}\right) ^{2}\left(
\mathcal{A}_{p}^{(0)}-\frac{1}{2}\mathcal{B}_{p}^{(0)}\right) \right] \text{
},
\end{multline}
where the Landau, $\mathcal{A}_{p}^{(r)}$, and Beliaev, $\mathcal{B}
_{p}^{(r)}$, coefficients $(r=0,1)$ are temperature independent (see\ Eqs.~(
\ref{BPi}) and (\ref{APi})). The asymptotic behaviors for $\gamma _{B}^{(p)}$
and $\gamma ^{(p)}$ are displayed by open squares in Figs.~\ref{fig:fig2}
and \ref{fig:fig5}, respectively. In the case of the Beliaev damping and by
comparison with the exact result in Fig.~\ref{fig:fig2}, we can see that the
thermal regime supported by Eq.~(\ref{HifT}), is a better approach for the
lower excited states. For example, for $p=2,$ Eq.~(\ref{HifT}) provides a
good result if $k_{B}T>0.5\hbar \omega _{0},$ while for $p=8,$ we have that
the function (\ref{HifT})\ is valid if $k_{B}T>1.3\hbar \omega _{0}.$ In
contrast, for the total damping, i.e. for higher values of $p$, the limit (
\ref{BETher}) becomes a better approach. Thus, for $p=2$ and 4, Eq.~(\ref
{BETher}) is good enough if $k_{B}T>\hbar \omega _{0}$ and $k_{B}T>0.6\hbar
\omega _{0},$ respectively. Notice that for $p=$ odd numbers, the thermal
regime (\ref{BETher}) is even better. These facts are explained by the
presence of the Landau matrix element, $\mathcal{A}_{p}^{(1)},$ in the total
damping calculation for $T\neq 0$ $K.$

The linear character of the damping rate with $T$ has been tested
experimentally in the atomic gas of Na.~\cite{Mewes} In the experiment of
Ref.~[\onlinecite{Mewes}] the condensate was loaded in a trap where the
transversal frequency $\omega _{r}>>\omega _{0}.$ Hence, we can argue that
we are in presence of a quasi-1D condensate. The excitation frequency
employed in the experiment was of $\omega _{ex}=1.58\omega _{0}$ and $\omega
_{0}=2\pi \times 19.3$ Hz. Following the Bogoliubov excitation spectrum of
Eq.~(\ref{frecue}), the excitation frequency $\omega _{ex}$ corresponds to
the $p=2$ confined phonon mode with a dimensional non-linear parameter $
\Lambda =$3.42. Using the asymptotic expression for the high temperature
regime, Eq.~(\ref{BETher}), we obtain $\gamma ^{(2)}=4.4$ $s^{-1}$ for $
T=200 $ nK and $17.6$ $s^{-1}$ for $T=800$ nK which agree quite well with
the reported experimental values of 4.4 $s^{-1}$ and 18 $s^{-1}$. In the
evaluation was assumed a condensate of 3500 atoms and from the value of $
\Lambda =$3.42 we extract an effective 1D coupling constant $g_{1}=3.7\times
10^{-25}$ eVm.

At very low temperature or strong confined regime, i.e. $k_{B}T<<\hbar
\omega _{0},$ from Eqs.~(\ref{LowT}) and (\ref{LaLowT}) follows that the
total damping of the exited mode $p$ can be cast as
\begin{equation}
\gamma ^{(p)}=\left[ \mathbb{A}_{p}^{\mathbb{(}1\mathbb{)}}(T)+\frac{1}{2}
\left( \mathbb{B}_{p}^{\mathbb{(}0\mathbb{)}}+\mathbb{B}_{p}^{\mathbb{(}1
\mathbb{)}}(T)\right) \right] ~.  \label{LimTlow}
\end{equation}
Here, the coefficients $\mathbb{A}_{p}^{\mathbb{(}1\mathbb{)}}$ and $\mathbb{
B}_{p}^{\mathbb{(}1\mathbb{)}}$ decay exponentially with $T$ and $\gamma
^{(p)}$ is almost constant independent of the temperature. Comparing the
results of Eq.~(\ref{LimTlow}) with the theoretical calculations for 3D or
2D homogeneous systems, where the trap potential and confined effect are
neglected, we found for the Landau damping a different behavior. Reference~[%
\onlinecite{ming}] reports the law $\gamma _{L}\sim T^{2}$, while the limit
of $\gamma _{L}\sim T^{4}$ is predicted in Ref.~[ %
\onlinecite{2Pitaevskii,Liu,Hohenberg}]. The quantum limit or very low
temperature for Believ damping and the total decay as a function of reduced
temperature, are represented by open diamonds in Figs.~\ref{fig:fig2} and %
\ref{fig:fig5}. From the figures it can be notice that, for $%
k_{B}T<<0.6\hbar \omega _{0}$, the asymptotic approach given by Eq.~(\ref%
{LimTlow}) reproduces quite well the decay processes.
\begin{figure}[tbp]
\includegraphics[width = 0.48\textwidth]{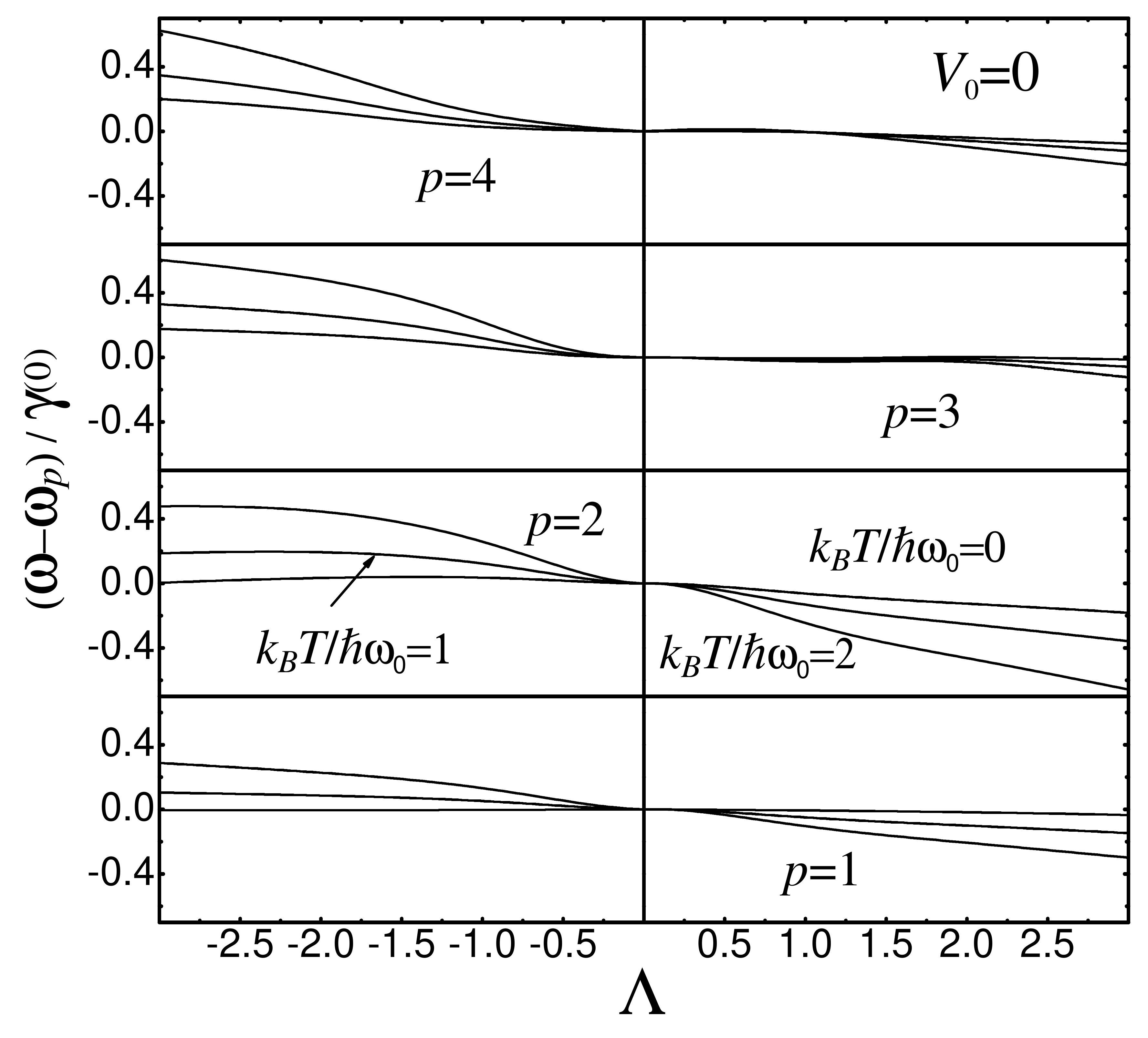}
\caption{Renormalized phonon frequencies, \emph{Re\{$\protect\pi _{p}$\}=$%
\Delta \widetilde{\protect\omega }_{p}$} in units of $\protect\gamma ^{(0)}$
as a function of the reduced self-interaction parameter $\Lambda $ for three
values of the temperature $k_{B}T/\hbar \protect\omega _{0}=0,1,2$ and
phonon states $p=1,2,3,4$.}
\label{fig:figRenor}
\end{figure}

An import result is the knowledge of the excited frequency shift as a
function of the condensate parameters and the applied laser intensity. The
real part of the self-energy in Eq.~(\ref{omega}) allows to an analytical
expression for the renormalized excited frequency, \emph{Re}\{$\pi _{p}$\}=$%
\Delta \widetilde{\omega }_{p},$ as a function of $\Lambda ,$ $V_{0}$ and $%
T. $ Thus,
\begin{multline}
\Delta \widetilde{\omega }_{p}=\gamma _{0}\left\vert \Lambda \right\vert
\sum_{i,j}\left[ \left( 1+f_{i}+f_{j}\right) \left\vert \overline{B_{ij}}%
\right\vert ^{2}\right.  \label{ren} \\
\frac{(\omega _{p}-\omega _{i}-\omega _{j})\omega _{0}}{(\omega _{p}-\omega
_{i}-\omega _{j})^{2}+\varepsilon ^{2}}+ \\
\left. \frac{1}{2}\left( f_{i}-f_{j}\right) \left\vert \overline{A_{ij}}%
\right\vert ^{2}\frac{(\omega _{j}-\omega _{i}-\omega _{p})\omega _{0}}{%
(\omega _{j}-\omega _{i}-\omega _{p})^{2}+\varepsilon ^{2}}\right] \text{ }.
\end{multline}%
Figures~(\ref{fig:figRenor}) shows the dependence of the dimensionless
self-interaction parameter $\Lambda $ on the renormalized discrete phonon
frequencies $\Delta \widetilde{\omega }_{p}=\omega -\omega _{p}$ for the
reduced temperature values $k_{B}T/\hslash \omega _{0}=0,$ $1$ and 2. We
conclude that for the attractive regime ($\Lambda <0$), the renormalized
shift $\Delta \widetilde{\omega }_{p}>0,$ while the opposite result is
obtained for the repulsive interaction, i.e. $\Delta \widetilde{\omega }%
_{p}<0$ if $\Lambda <0.$ This behavior is understood by the dependence of $%
\Delta \widetilde{\omega }_{p}$ in Eq.~(\ref{ren}) on the label spacing $%
\Delta _{p}^{(i,j)}=\left( \omega _{p}-\omega _{i}-\omega _{j}\right)
/\omega _{0}$ and $\Delta _{j}^{(i,p)}=\left( \omega _{j}-\omega _{i}-\omega
_{p}\right) /\omega _{0}$ as a function of $\Lambda .$ According to the
results of Appendixes A anb B, we have that in the thermal regime, $\Delta
\widetilde{\omega }_{p}$ is proportional to $k_{B}T/\hslash \omega _{0}$.
Thus a linear increase or decrease of the excited frequency with the
temperature is predicted for attractive or repulsive interaction between
atoms, respectively.

In conclusion, we evaluated the damping rates of confined phonon modes of 1D
condensates in a harmonic trap potential loaded in an optical lattice. We
remarked the influence of the spatial confinement potential on the
collective oscillations and on the damping rates as a function of the
temperature. The presence of an optical lattice as an external field, allows
to manipulate the decay rate of the condensate. The damping $\gamma ^{(p)}$
can be turned on or turned off as a function on of the laser intensity.
Also, for a given excited state $p$ and tuning the laser intensity, it is
possible to get a set of transitions, $\omega _{p}\rightarrow \omega _{j}\pm
\omega _{i},$ reaching to a \textit{\ resonant effect} for the total
lifetime $1/\gamma ^{(p)}.$

\section*{Acknowledgments}

The authors acknowledge financial support of Brazilian agencies CNPq and
FAPESP. Part of the work has been supported by Alexander von Humboldt
Foundation. C.-T.-G. is grateful to the Max-Planck-Institute for Complex
Systems and the Universidade Federal do Rio de Janeiro for their hospitality.

\appendix

\section{Beliaev matrix element}

After substituting the perturbed wave function $\left\vert
u_{p}\right\rangle $ and $\left\vert v_{p}\right\rangle $ in the matrix
element (\ref{Bij}), and neglecting terms of the second order or higher in $%
\Lambda $ and $V_{0},$ for the function $\mathcal{M}_{p}^{(B)}$ we get%
\begin{equation}
\mathcal{M}_{p}^{(B)}(\Lambda ,V_{0})=\sum_{i,j}\left( 1+f_{i}+f_{j}\right)
\left\vert \overline{B_{ij}}\right\vert ^{2}\mathcal{L}_{p}^{(+)}\text{ },
\label{Bel}
\end{equation}%
where $\overline{B_{ij}}=T_{0pij}-\Lambda F_{pij}+V_{0}H_{pij}$, $T_{lpij}$
( $l+p+i+j=$ even number) is reported elsewhere.~\cite{trallero2}
\begin{equation}
F_{pij}=a_{pij}+b_{jpi}^{+}+b_{ijp}^{+}+2(b_{pij}^{-}+b_{ijp}^{-}+b_{jpi}^{-})%
\text{ ,}
\end{equation}%
\begin{equation}
H_{pij}=c_{pij}+d_{pij}+d_{ijp}+d_{jpi}\text{ ,}
\end{equation}%
\begin{eqnarray}
a_{pij} &=&\sum_{m\neq 0}\frac{T_{2m000}T_{2mpij}}{2m}\text{ };\text{ \ \ }%
b_{pij}^{\pm }=\sum_{m}^{\prime }\frac{T_{00pm}T_{0mij}}{m\pm p}\text{ },
\notag \\
c_{pij} &=&\sum_{m\neq 0}\frac{g_{0,2m}T_{0mij}}{2m}\text{ };\text{ \ \ }%
d_{pij}=\sum_{m\neq p}\frac{g_{p,m}T_{0mij}}{2(m-p)}\text{ ,}  \notag \\
&&
\end{eqnarray}%
with the parity condition\ $p+i+j=$ even number, and the Lorenzian function
\begin{equation*}
\mathcal{L}_{p}^{(\pm )}=\frac{1}{\pi}\frac{\omega _{0}\varepsilon }{(\omega
_{p}\mp \omega _{i}-\omega _{j})^{2}+\varepsilon ^{2}}\text{ .}
\end{equation*}%
In the limit of thermal regime, the probability $\mathcal{M}_{p}^{(B)}$ is
reduced to%
\begin{equation}
\mathcal{M}_{p}^{(B)}(T)=\frac{k_{B}T}{\hbar \omega _{0}}\left[ \mathcal{B}%
_{p}^{(1)}+\frac{1}{12}\left( \frac{\hbar \omega _{0}}{k_{B}T}\right) ^{2}%
\mathcal{B}_{p}^{(0)}\right] \text{ },  \label{HifT}
\end{equation}%
with%
\begin{equation}
\mathcal{B}_{p}^{(r)}=\sum_{i,j}\frac{\omega _{i}+\omega _{j}}{\omega _{0}}
\left( \frac{\omega _{0}^{2}}{\omega _{i}\omega _{j}}\right) ^{r}\left\vert
\overline{B_{ij}}\right\vert ^{2}\mathcal{L}_{p}^{(+)}\text{ ; \ }\ (r=0,1)%
\text{ }.  \label{BPi}
\end{equation}%
For low temperature we have
\begin{equation}
\mathcal{M}_{p}^{(B)}(T)=\mathbb{B}_{p}^{\mathbb{(}0\mathbb{)}}+\mathbb{B}%
_{p}^{\mathbb{(}1\mathbb{)}}(T)\text{ }.  \label{LowT}
\end{equation}%
where%
\begin{multline}
\mathbb{B}_{p}^{\mathbb{(}r\mathbb{)}}=\sum_{i,j}\left( \exp (-\hbar \omega
_{i}/k_{B}T)+\exp (-\hbar \omega _{j}/k_{B}T)\right) ^{r}  \label{BLowT} \\
\times \left\vert \overline{B_{ij}}\right\vert ^{2}\mathcal{L}_{p}^{(+)}%
\text{ ; \ }\ (r=0,1)\text{ }.
\end{multline}

\section{Landau{\ matrix element}}

Using the wave function $\left\vert u_{p}\right\rangle $, $\left\vert
v_{p}\right\rangle $ and Eq.~(\ref{A}) and neglecting terms higher than $%
\Lambda $ and $V_{0}$ we have for $\mathcal{M}_{p}^{(B)}$%
\begin{equation}
\mathcal{M}_{p}^{(L)}=\sum_{i,j}\left( f_{i}-f_{j}\right) \left\vert
\overline{A_{ij}}\right\vert ^{2}\mathcal{L}_{p}^{(-)}\text{ },  \label{Land}
\end{equation}%
where $\overline{A_{ij}}=T_{0pij}-\Lambda D_{pij}+V_{0}G_{pij}$,%
\begin{eqnarray*}
D_{pij}
&=&a_{pij}+b_{pij}^{+}+b_{ijp}^{+}+2(b_{pij}^{-}+b_{ijp}^{-}+b_{jpi}^{-})%
\text{ }, \\
G_{pij} &=&c_{pij}+d_{pij}+d_{ijp}+d_{jpi}\text{ }.
\end{eqnarray*}%
If $k_{B}T<<\hbar \omega _{0}$, the Landau probability process can be
approached to
\begin{equation}
\mathcal{M}_{p}^{(L)}(T)=\frac{k_{B}T}{\hbar \omega _{0}}\left[ \mathcal{A}%
_{p}^{(1)}-\frac{1}{12}\left( \frac{\hbar \omega _{0}}{k_{B}T}\right) ^{2}%
\mathcal{A}_{p}^{(0)}\right]
\end{equation}%
with%
\begin{equation}
\mathcal{A}_{p}^{(r)}=\sum_{i,j}\frac{\omega _{j}-\omega _{i}}{\omega _{0}}
\left( \frac{\omega _{0}^{2}}{\omega _{i}\omega _{j}}\right) ^{r}\left\vert
\overline{A_{ij}}\right\vert ^{2}\mathcal{L}_{p}^{(-)}\text{ };\ \ (i=0,1)
\text{ }.  \label{APi}
\end{equation}%
While, for the weak confinement, $k_{B}T>>\hbar \omega _{0},$ it is obtained
that
\begin{multline}
\mathcal{M}_{p}^{(L)}(T)=\mathbb{A}_{p}^{\mathbb{(}1\mathbb{)}}=\sum_{i,j}%
\left[ \exp (-\hbar \omega _{i}/k_{B}T)\right.  \label{LaLowT} \\
\left. -\exp (-\hbar \omega _{j}/k_{B}T)\right] \\
\times \left\vert \overline{A_{ij}}\right\vert ^{2}\mathcal{L}_{p}^{(-)}%
\text{ }.
\end{multline}

\end{document}